\documentclass{article}

\usepackage{siunitx}
\usepackage{amsmath}
\usepackage{amssymb}
\usepackage{amsfonts}
\usepackage{epsfig}
\usepackage{graphicx}
\usepackage[nomarkers]{endfloat}
\usepackage{color}
\definecolor{cor}{rgb}{1,0,0} 
\usepackage{upgreek}
\usepackage{bm}

\usepackage{epsfig}
\usepackage{graphics}
\usepackage{rotating}
\usepackage{natbib}
\usepackage[section] {placeins}

\usepackage[T3,T1]{fontenc}
\DeclareSymbolFont{tipa}{T3}{cmr}{m}{n}
\DeclareMathAccent{\invbreve}{\mathalpha}{tipa}{16}

\begin{document}
	
	\title{3D Marchenko internal multiple attenuation on narrow azimuth streamer data of the Santos Basin, Brazil}
	
	\begin{center}
		\rule{0cm}{2cm}
		{\bf \LARGE 3D Marchenko internal multiple attenuation on narrow azimuth streamer data of the Santos Basin, Brazil}\\
		\rule{0cm}{1cm}\\
		Myrna Staring and Kees Wapenaar
	\end{center}

\begin{abstract}
	
	In recent years, a variety of Marchenko methods for the attenuation of internal multiples has been developed. These methods have been extensively tested on 2D synthetic data and applied to 2D field data, but only little is known about their behaviour on 3D synthetic data and 3D field data. Particularly, it is not known whether Marchenko methods are sufficiently robust for sparse acquisition geometries that are found in practice. Therefore, we start by performing a series of synthetic tests to identify the key acquisition parameters and limitations that affect the result of 3D Marchenko internal multiple prediction and subtraction using an adaptive double-focusing method. Based on these tests, we define an interpolation strategy and use it for the field data application.
	
	Starting from a wide azimuth dense grid of sources and receivers, a series of decimation tests is performed until a narrow azimuth streamer geometry remains. We evaluate the effect of the removal of sail lines, near offsets, far offsets and outer cables on the result of the adaptive double-focusing method. These tests show that our method is most sensitive to the limited aperture in the crossline direction and the sail line spacing when applying it to synthetic narrow azimuth streamer data. The sail line spacing can be interpolated, but the aperture in the crossline direction is a limitation of the acquisition.  
	
	Next, we apply the adaptive Marchenko double-focusing method to the narrow azimuth streamer field data from the Santos Basin, Brazil. Internal multiples are predicted and adaptively subtracted, thereby improving the geological interpretation of the target area. These results imply that our adaptive double-focusing method is sufficiently robust for the application to 3D field data, although the key acquisition parameters and limitations will naturally differ in other geological settings and for other types of acquisition.

\end{abstract}

\section{Introduction}

\begin{figure}[bt]
	\centering
	\includegraphics[width=8cm]{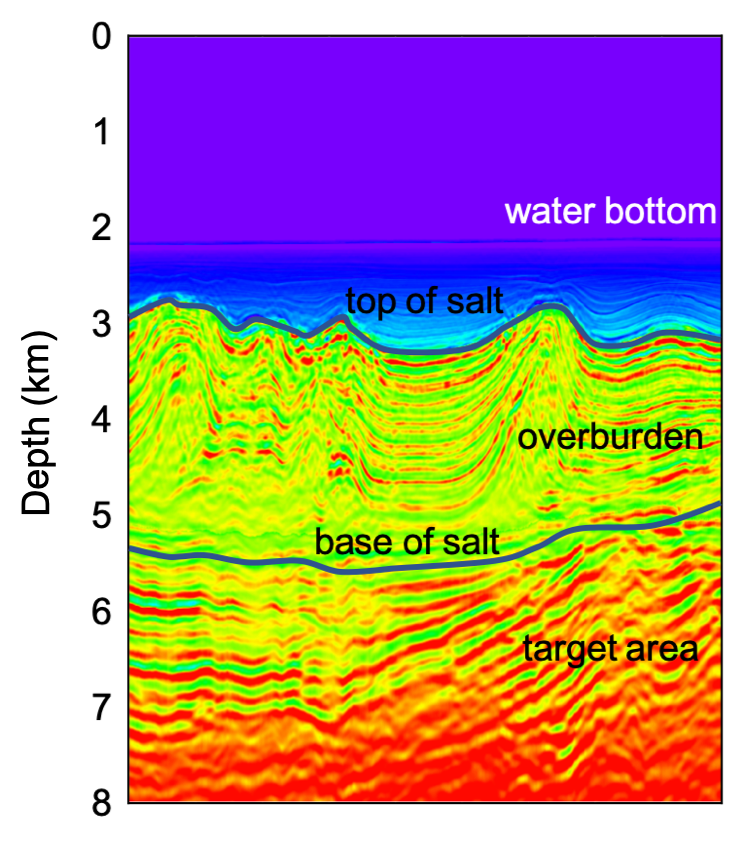}
	\caption{A 2D slice of the velocity model of the Santos Basin, Brazil.
		\label{fig:vel_mod}}
\end{figure}

The Santos Basin in Brazil is known for its oil-bearing carbonate reservoirs below a highly reflective stratified salt layer (see figure \ref{fig:vel_mod}). The salt layer generates strong internal multiples that pose a problem for seismic imaging \citep{cypriano2015impact}. Most imaging methods assume that the recorded wavefield was only reflected once and thus incorrectly interpret internal multiples as primaries from deeper reflectors. As a result, these methods create ghost reflectors that do not exist in reality. These ghost reflectors can interfere with the real reflectors in the target area and thereby corrupt the image. Therefore, we wish to attenuate internal multiples in order to obtain a reliable image of the target area. 

The attenuation of internal multiples is a challenge. Various methods have been proposed, ranging from filtering methods \citep[e.g.][]{hampson1986inverse, foster1992suppression, zhou1994wave} that transform the reflection response to an alternative domain in which the primaries and the internal multiples separate, to wave equation-based methods that aim to predict the internal multiples by convolving and correlating the reflection response with itself \citep[e.g.][]{jakubowicz1998wave, weglein1997inverse}. The application of filtering methods is often challenging in settings with a complex overburden, since there is usually no distinct difference in properties between the primary reflections from the target area and the (strong) internal multiples generated in an overburden. Also, the application of wave equation-based methods is not undisputed. Some wave equation-based methods require the manual identification of internal multiple generators, thereby introducing bias and the risk of not correctly capturing all internal multiple generators in the process. In addition, some wave equation-based methods predict internal multiples with incorrect amplitudes or use a layer stripping approach that results in error accumulation from shallow to deep reflectors. In order to attenuate internal multiples in a complex setting such as the Santos Basin, an alternative method is needed. 

Marchenko methods \citep{ware1969continuous, broggini2012focusing, wapenaar2014marchenko} are data-driven and wave-equation-based methods that do not have these drawbacks. These methods have the ability to consider the entire overburden as a whole, instead of having to identify all individual internal multiple generators separately. In addition, Marchenko methods allow us to retrieve Green's functions including primaries as well as all orders of internal multiples at any desired depth level without having to resolve overlying layers first. When writing the retrieval of Green's functions using the coupled Marchenko equations as a Neumann series, Marchenko methods can be used for the prediction of internal multiples \citep{van2015illustration}. These predictions in principle have the correct amplitude and phase. However, minor amplitude and phase differences are usually present when applying the method to field data due to imperfect acquisition or preprocessing. A mild adaptive filter can be used to correct for these minor differences. We previously reported on the successful application of an adaptive Marchenko method (the adaptive double-focusing method) to 2D synthetic data and a 2D line of streamer data of the Santos Basin in Brazil \citep{staring2018source}. Internal multiples were predicted and adaptively subtracted from the target area, which improved the geological interpretation. In addition, we found that the adaptive double-focusing method was relatively robust for a sparse acquisition geometry in 2D and suitable for the application to large data volumes. In the hope that these properties also hold in 3D, we use this adaptive Marchenko method for the prediction and adaptive subtraction of internal multiples from 3D narrow azimuth streamer data acquired in the Santos Basin. 

The extension from 2D to 3D Marchenko methods may seem trivial in theory, but it is not the case in practice. Some aspects are similar, such as the data preparation requirements that include noise suppression, signature deconvolution, deghosting and the removal of surface-related multiples. However, aspects related to the sampling of the acquired data are different. In addition to the inline direction in 2D, there is in 3D also a crossline direction that typically has a limited aperture and less densely spaced sources and receivers. Also, streamers usually do not record responses at negative offsets, near offsets and far offsets in the inline direction. A thorough understanding of the effect of these acquisition limitations on the result of Marchenko internal multiple attenuation would allow us to estimate whether the application to any particular dataset is feasible. In addition, it would aid us in defining an interpolation strategy. Eventhough some researchers already applied a Marchenko method to 3D field data \citep{staring2018marchenko, pereira2018efficient}, they did not address the acquisition requirements and limitations of 3D Marchenko methods in detail. The objective of this paper is to gain a better understanding of the key acquisition parameters and limitations that affect the application of the adaptive double-focusing method to 3D data.

\begin{figure}
	\centering
	\includegraphics[width=14cm]{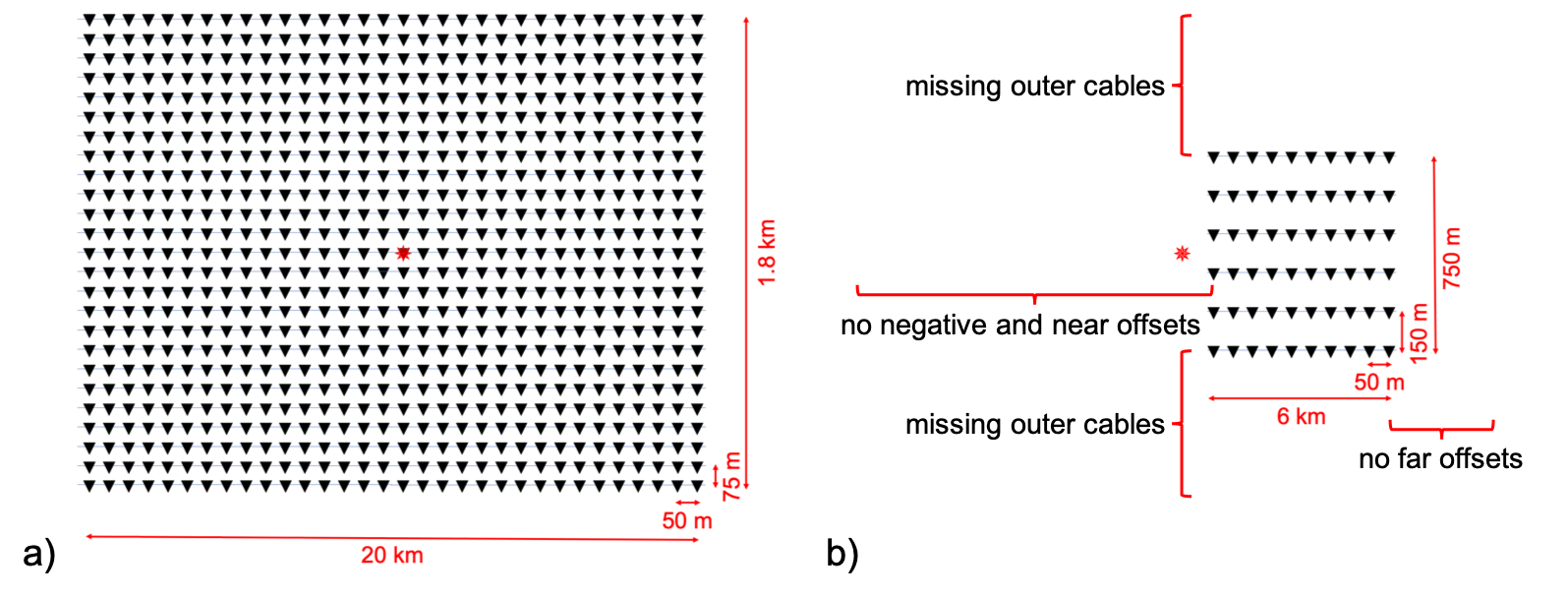}
	\caption{Cartoons showing a) the starting acquisition geometry and b) the final acquisition geometry for the synthetic decimation tests in this paper. The final acquisition geometry is based on our narrow azimuth streamer data. The stars represent sources and the triangles represent receivers. 
		\label{fig:grids}}
\end{figure}

In this paper, we first revise the theory of the adaptive Marchenko double-focusing method. Second, we perform a series of 3D synthetic tests to study the effect of the acquisition parameters on the result of internal multiple prediction and adaptive subtraction using this method. Starting from a grid spacing of 50 m (inline direction) by 75 m (crossline direction) co-located sources and receivers with positive and negative offsets, near offsets, far offsets and a crossline aperture of 1.8 km (figure \ref{fig:grids}a), we step-by-step decimate the acquisition down to a narrow azimuth streamer geometry on which our 3D field data were acquired (figure \ref{fig:grids}b). Based on these tests, we identify the key limiting acquisition parameters and use these to design an interpolation strategy for the field data application. Next, we test the proposed interpolation strategy on 3D synthetic data. Finally, we apply the adaptive double-focusing method to 3D narrow azimuth streamer data. In the following discussion and conclusion section, we evaluate the performance of the adaptive double-focusing method.

\section{Marchenko internal multiple attenuation by adaptive double-focusing}

The adaptive Marchenko double-focusing method requires a preprocessed reflection response $R (\textbf{x}_R,\textbf{x}_S,t)$ acquired on a sufficiently dense grid of sources $\textbf{x}_S$ and receivers $\textbf{x}_R$ at the acquisition surface $\partial \mathbb{D}_0$. A smooth velocity model of the subsurface is needed to obtain the direct wave of the downgoing focusing function $\invbreve{f}_0^+$. The direct wave is obtained by modeling and time-reversing the response from sources at the redatuming level $\partial \mathbb{D}_i$ to receivers at the acquisition surface $\partial \mathbb{D}_0$ using finite-difference modeling or an Eikonal solver (see figure \ref{fig:direct}). The $\ \invbreve{\cdot}\ $ symbol indicates an user-specified wavelet that is convolved with the modeled wavefield. The direct downgoing focusing function $\invbreve{f}_0^+$ initiates the iterative scheme that solves the coupled Marchenko equations. If the overburden were homogeneous, this initial wavefield would be sufficient to create a focus at the desired focal point at $\partial \mathbb{D}_i$. Otherwise, a coda for the downgoing focusing function has to be retrieved using the following series \citep{van2015green}:

\begin{figure}
	\centering
	\includegraphics[width=4cm]{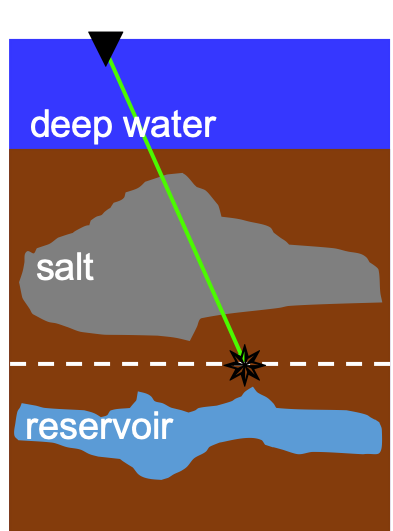}
	\caption{Cartoon illustrating the direct wave $\invbreve{f}_0^+$, which we obtain using a smooth velocity model and an Eikonal solver.
		\label{fig:direct}}
\end{figure} 

\begin{equation}
\label{f+}
\invbreve{f}^{+} (\textbf{x}_S, \textbf{x}_F', t )
=
\sum_{i=0}^\infty
\invbreve{f}^{+}_{i} (\textbf{x}_S, \textbf{x}_F', t )
=
\sum_{i=0}^\infty
\{\Theta \mathcal{R}^{\star}  \Theta \mathcal{R}\}^i
\invbreve{f}_{0}^+ (\textbf{x}_S, \textbf{x}_F', t),
\end{equation}

where $i$ is the iteration number. Symbol $\textbf{x}_F'$ denotes focal points at the redatuming level $\partial \mathbb{D}_i$ that become virtual sources. Operators $\mathcal{R}$ and $\mathcal{R}^{\star}$ perform a multidimensional convolution or correlation of the reflection response $R$ with the wavefield that it acts upon. Window functions $\Theta$ are tapered Heaviside step functions that separate the causal and the acausal wavefields (i.e. Green's functions and focusing functions) in time. See appendix A of \cite{staring2018source} for details on the design of window function $\Theta$. The first update of the coda of the downgoing focusing function $\invbreve{f}_1^+$ already contains many of the correct events to compensate for the inhomogeneous overburden, but with incorrect amplitude. Higher-order estimates ($i=2,3,4,etc.$) are needed to obtain the correct amplitude.  

Using the downgoing focusing function $\invbreve{f}^{+}$, we can also retrieve the receiver redatumed upgoing Green's function:

\begin{equation}
\label{g-}
\begin{split}
\invbreve{G}^{-}  (\textbf{x}_{F}, \textbf{x}_S, t)
&=
\Psi
\mathcal{R}
\invbreve{f}^{+} (\textbf{x}_S, \textbf{x}_{F}, t )\\
&=
\sum_{j=0}^\infty
\invbreve{G}^{-}_j  (\textbf{x}_{F}, \textbf{x}_S, t)
=
\Psi
\mathcal{R}
\sum_{j=0}^\infty
\{\Theta \mathcal{R}^{\star}  \Theta \mathcal{R}\}^j
\invbreve{f}_{0}^+ (\textbf{x}_S, \textbf{x}_{F}, t),\\
\end{split}
\end{equation}

where mute $\Psi=I-\Theta$ now selects the causal wavefield and symbol $\textbf{x}_F$ represents focal points at the redatuming level $\partial \mathbb{D}_i$ that become virtual receivers. The iteration number is given by $j$. Initial estimate $\invbreve{G}_0^{-}$ is the standard receiver-redatumed upgoing Green's function at $\textbf{x}_{F}$. The first update $\invbreve{G}_1^{-}$ contains a first-order estimate of the receiver-side internal multiples generated in the overburden with incorrect amplitude. Next updates ($\invbreve{G}_2^{-}$, $\invbreve{G}_3^{-}$, etc.) contain higher-order estimates that are necessary to obtain the correct amplitude. An additional step is needed to also remove source-side and source-and-receiver-side internal multiples generated by the overburden.

The retrieval of the upgoing Green's function $\invbreve{G}^{-}$ with a grid of sources at the acquisition surface $\partial \mathbb{D}_0$ and a grid of virtual receivers at the redatuming level $\partial \mathbb{D}_i$ is a single-focusing step. By creating double-focusing we also remove other internal multiples generated by the overburden. To this end, we convolve the upgoing Green's function $\invbreve{G}^{-}$ at virtual receivers with the downgoing focusing function $\invbreve{f}^{+}$ at virtual sources \citep{wapenaar2016unified, van2018single, staring2018source}:

\begin{equation}
\label{double}
\invbreve{\invbreve{G}}^{-+} (\textbf{x}_{F},\textbf{x}_F',  t )=\int_{\partial \mathbb{D}_0}  \invbreve{G}^- (\textbf{x}_{F}, \textbf{x}_S, t) *   \invbreve{f}^+  (\textbf{x}_S, \textbf{x}_F', t) d^2 \textbf{x}_S.
\end{equation}

\begin{figure}
	\centering
	\includegraphics[width=12cm]{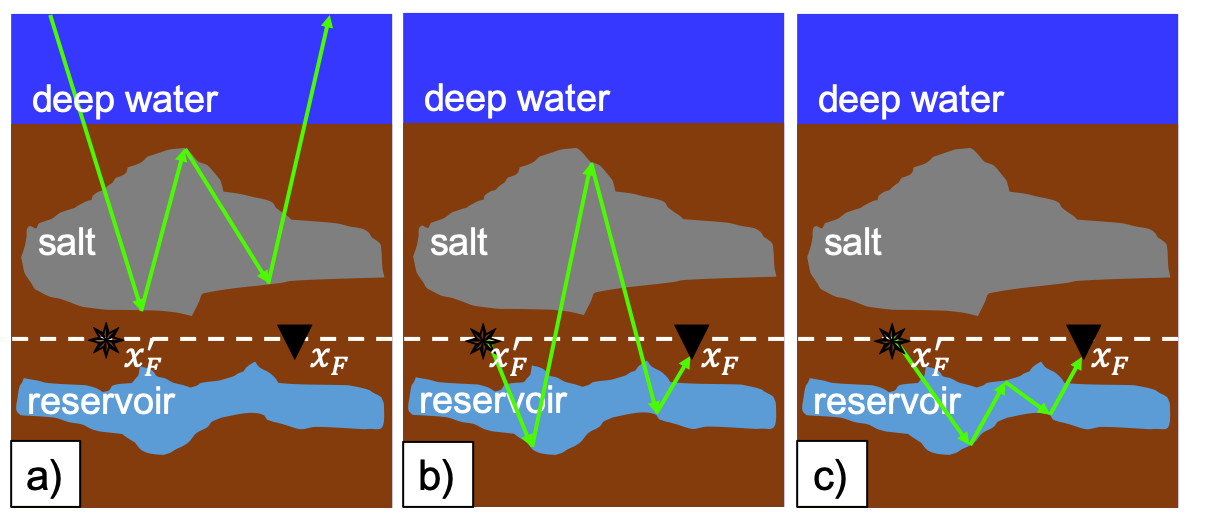}
	\caption{Cartoons illustrating the redatumed sources at $\textbf{x}_F'$ and the redatumed receivers at $\textbf{x}_F$ resulting from the adaptive double-focusing method. Internal multiples a) generated by the overburden have been removed by double-focusing, while b) later arriving internal multiples generated by interactions between the target area and the overburden and c) later arriving internal multiples generated by the target area remain. 
		\label{fig:multiples}}
\end{figure}

By applying this for many positions $\textbf{x}_{F}$ and $\textbf{x}_F'$ at redatuming level $\partial \mathbb{D}_i$, a grid of downward radiating virtual sources and virtual receivers that measure the upgoing wavefield is created. The result is a redatumed Green's function $\invbreve{\invbreve{G}}^{-+}$ in the physical medium. Internal multiples generated by the overburden (figure \ref{fig:multiples}a) have been removed, but later arriving internal multiples generated by interactions between the target area and the overburden (figure \ref{fig:multiples}b) and internal multiples generated by the target area (figure \ref{fig:multiples}c) remain. According to \cite{cypriano2015impact}, the main internal multiples that contaminate the image of the target area in the Santos Basin are generated between the water bottom and the top of salt (see figure \ref{fig:vel_mod}). By using double-focusing, we remove these internal multiples, while leaving some internal multiples below the target area behind. Note that one user-specified wavelet $\ \invbreve{\cdot} \ $  has to be deconvolved from the redatumed Green's function $\invbreve{\invbreve{G}}^{-+}$. Also note that the integral over the acquisition surface $\partial \mathbb{D}_0$ allows us to parallelize the implementation of the double-focusing method per pair of focal points, which makes this method particularly suitable for the application to large 3D data volumes.

Using equations \ref{f+} and \ref{g-}, we can write equation \ref{double} as a series:

\begin{equation} \label{eq:double2} 
\begin{split}
&
\qquad \qquad
\invbreve{\invbreve{G}}^{-+}
(\textbf{x}_{F},\textbf{x}_F',  t ) =\\
&\sum_{j=0}^\infty
\sum_{i=0}^\infty
\int_{\partial \mathbb{D}_0}
{\invbreve{G}}_j^{-} (\textbf{x}_{F},\textbf{x}_S,t) *
{\invbreve{f}}_{i}^{+} 	(\textbf{x}_S,\textbf{x}_F', t)
d^2 {\textbf{x}_S}\\
&\approx
\int_{\partial \mathbb{D}_0}
{\invbreve{G}}_0^{-} (\textbf{x}_{F},\textbf{x}_S,t) *
{\invbreve{f}}_0^{+} (\textbf{x}_S,\textbf{x}_F',t)
d^2 {\textbf{x}_S}\\
&-
\int_{\partial \mathbb{D}_0} -
{\invbreve{G}}_1^{-} (\textbf{x}_{F},\textbf{x}_S,t) *
{\invbreve{f}}_0^{+} (\textbf{x}_S,\textbf{x}_F',t)
d^2 {\textbf{x}_S}\\
&-
\int_{\partial \mathbb{D}_0} -
{\invbreve{G}}_0^{-} (\textbf{x}_{F},\textbf{x}_S,t) *
{\invbreve{f}}_{1}^{+} 	(\textbf{x}_S,\textbf{x}_F',t)
d^2 {\textbf{x}_S}\\
&-
\int_{\partial \mathbb{D}_0} -
{\invbreve{G}}_1^{-} (\textbf{x}_{F},\textbf{x}_S,t) *
{\invbreve{f}}_{1}^{+} 	(\textbf{x}_S,\textbf{x}_F',t)
d^2 {\textbf{x}_S}\\
& -...
\end{split}
\end{equation}

The first term $\invbreve{G}_0^{-} * \invbreve{f}_0^{+}$ is the standard source and receiver redatumed Green's function including primaries and internal multiples. The second term $- \invbreve{G}_1^{-} * \invbreve{f}_{0}^{+}$ contains first-order predictions of receiver-side internal multiples generated by the overburden, while the third term $- \invbreve{G}_0^{-} * \invbreve{f}_{1}^{+}$ contains first-order predictions of source-side internal multiples generated by the overburden and the fourth term $- \invbreve{G}_1^{-} * \invbreve{f}_{1}^{+}$ contains first-order predictions of source-and-receiver-side internal multiples generated by the overburden. Subsequent terms contain higher-order estimates of the predicted internal multiples that are needed to obtain the correct amplitude. Note that Marchenko methods in principle do not rely on an adaptive filter to accurately attenuate internal multiples. However, instead of retrieving all terms in the series in equation \ref{eq:double2} by correlating and convolving the data with itself many times (see equation \ref{g-}), we propose to retrieve only a few updates and use an adaptive filter as a substitute for higher-order amplitude corrections. Also, an adaptive filter can correct for any minor amplitude and phase differences that are present in the internal multiple predictions due to imperfections in the data acquisition or preprocessing. Iteration numbers $i$ and $j$ that are needed to obtain predictions of all overburden internal multiples depend on the geological setting. In our case, new internal multiples were not predicted beyond the third term in equation \ref{eq:double2}, so we only use the terms $- \invbreve{G}_1^{-} * \invbreve{f}_{0}^{+}$ and $- \invbreve{G}_0^{-} * \invbreve{f}_{1}^{+}$ for the prediction of internal multiples in this particular setting. These predictions are treated as individual internal multiple predictions, which are orthogonalized to the data prior to simultaneous adaptive subtraction. We have chosen for an adaptive filter in the curvelet domain \citep{herrmann2008adaptive, wu2015high}, since it can distinguish between primaries and internal multiples in space, time and dip. Naturally, care has to be taken not to subtract the primary reflections with the internal multiples. 

\section{Sensitivity tests on 3D synthetic data}

We perform a series of 3D synthetic tests to identify the key acquisition parameters that affect the result of the adaptive double-focusing method. In order to generate synthetic data that represent the geological contrasts in the area as realistically as possible, we use a velocity model (see a 2D slice in figure \ref{fig:vel_mod}) and a density model that are obtained from an acoustic inversion of field data based on the original seismic image and migration velocity. The grid size of these models is 18.75 m by 18.75 m by 10 m. Co-located sources and receivers are positioned with a spacing of 50 m in the inline direction and a spacing of 75 m in the crossline direction (figure \ref{fig:grids}a), thereby simulating an inline spacing of 50 m and a sail line spacing of 75 m. The inline aperture is 20 km (offsets from -10 km to 10 km) and the crossline aperture is 1.8 km. An acoustic finite-difference algorithm is used to model data up to 30 Hz, such that the dominant wavelength at the receivers is 50 m. The recording time is 8.5 s. Also, we generate an initial focusing function $\invbreve{f}_{0}^{+}$ in the smooth velocity model using an Eikonal solver. Geometrical spreading  is part of the simulation. In addition, we convolve the response with an Ormsby wavelet with tapers at the low and the high ends. 

Starting from 24 lines of data modeled on this dense acquisition grid, we step-by-step decimate down to a realistic streamer acquisition geometry with a cable spacing of 150 m, a sail line spacing of 450 m and a cable length of 6 km. The inline source and receiver spacing remain 50 m. Inline offsets range from 250 m to 6250 m and the crossline aperture is 0.75 km (figure \ref{fig:grids}b). Throughout the decimation tests, we use the Marchenko double-focusing method to redatum to a grid of co-located virtual sources and virtual receivers below the overburden with a spacing of 25 m by 37.5 m. We have chosen the redatuming level to be just above the base of salt. The base of salt is the top of our reservoir and is therefore part of the target area. The main internal multiple generators in this geological setting, the water bottom and the top of salt, are part of the overburden. Internal multiple predictions are obtained by convolving the individual updates of the wavefields $\invbreve{G}_j^{-}$ and $\invbreve{f}_i^{+}$, which are subsequently subtracted from the data using a 3D curvelet filter \citep{wu2015high}. We orthogonalize the predictions and the data before subtraction, but do not use a global least squares filter for pre-conditioning. Parameters that need to be set are the number of scales in the transform, the number of angles in the transform, the window size and some sparsity parameters that control the inversion. We extensively test different filter settings and obtain the best results (the least damage of the primary reflections) using 7 scales, 8 angles and tapered windows of 768 ms by 256 traces. These settings are used for all synthetic examples shown here.

\subsection{The complete data set}

\begin{figure}
	\centering
	\includegraphics[width=10cm]{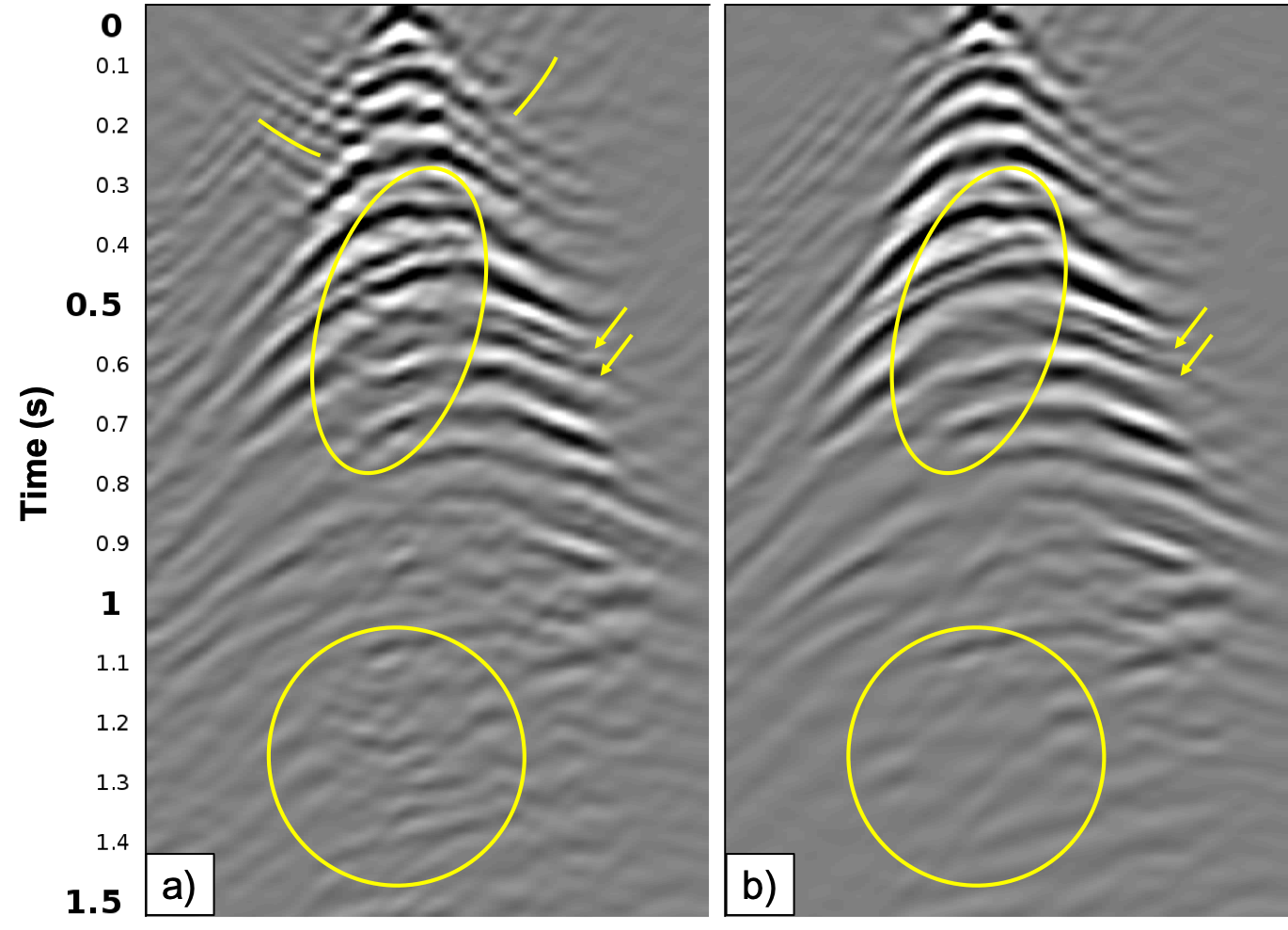}
	\caption{Redatumed common source gathers showing the result of the adaptive double-focusing method applied to 3D synthetic data modeled on the dense acquisition grid in figure \protect\ref{fig:grids}a. The gathers show: a) the redatumed Green's function $G_0^{-} * \invbreve{f}_{0}^{+}$ including primaries and internal multiples and b) the redatumed Green's function $\invbreve{G}^{-+}$ after prediction and adaptive subtraction of internal multiples. The yellow ellipses, stripes and arrows indicate areas in which internal multiple attenuation is most visible.
		\label{fig:50}}
\end{figure}

First, we apply the adaptive double-focusing method to synthetic data generated on the dense grid in figure \ref{fig:grids}a. Figure \ref{fig:50} shows redatumed common source gathers before and after internal multiple prediction and subtraction ($\invbreve{G}_0^{-} * \invbreve{f}_{0}^{+}$ and $\invbreve{\invbreve{G}}^{-+}$ from equation \ref{eq:double2}). The common source gathers are from a virtual source in the middle of the grid of focal points, as indicated by the red star in figure \ref{fig:grids}a. A difference is visible, especially in the yellow ellipses and along the yellow lines. It seems that conflicting seismic events were resolved, resulting in a better continuity of the primary events. 

\begin{figure}
	\centering
	\includegraphics[width=14cm]{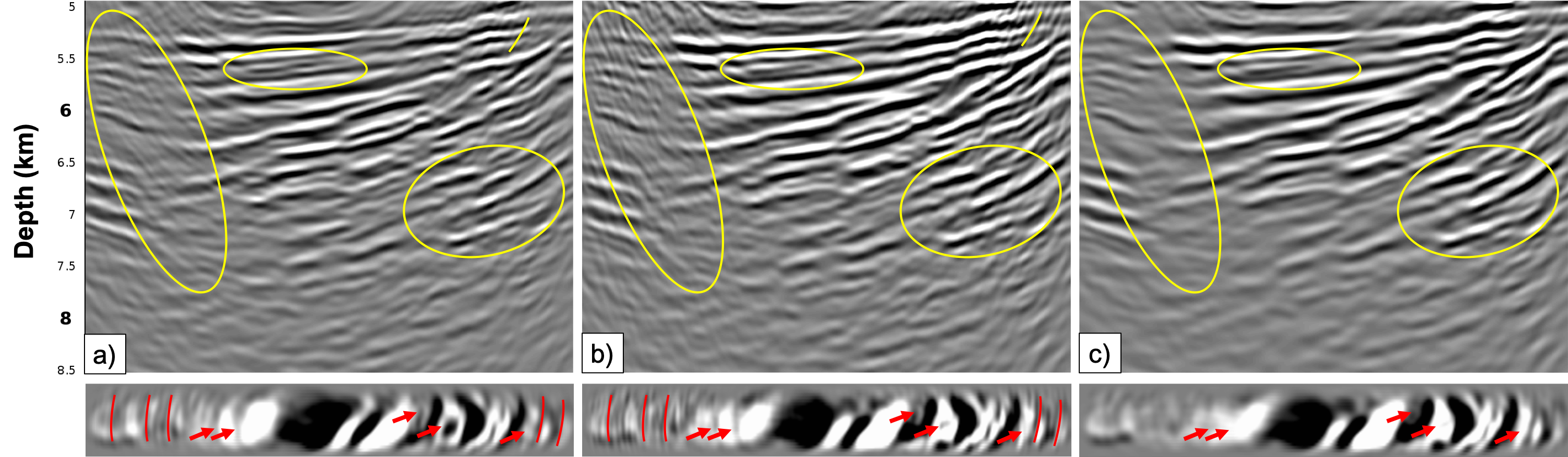}
	\caption{RTM images showing the result of the adaptive double-focusing method applied to 3D synthetic data modeled on the dense acquisition grid in figure \protect\ref{fig:grids}a. The images show: a) the migrated reflection response $\invbreve{R}$, b) the redatumed and migrated Green's function $G_0^{-} * \invbreve{f}_{0}^{+}$ including primaries and internal multiples, and c) the redatumed and migrated Green's function $\invbreve{G}^{-+}$ after prediction and adaptive subtraction of internal multiples. Below the RTM images are depth slices at 5900 m. The yellow ellipses indicate areas in which internal multiple attenuation is most visible.
		\label{fig:50_rtm}}
\end{figure}

Next, we deconvolve an user-specified wavelet $\ \invbreve{\cdot} \ $ and migrate the result. Figure \ref{fig:50_rtm} shows RTM images of the reflection response $\invbreve{R}$ at the acquisition surface (note that this image was truncated at the base of salt for comparison), the redatumed Green's function including primaries and internal multiples $G_0^{-} * \invbreve{f}_{0}^{+}$ and the redatumed Green's function $\invbreve{G}^{-+}$ after internal multiple prediction and subtraction. Figures \ref{fig:50_rtm}a and \ref{fig:50_rtm}b are comparable, thereby demonstrating that source-receiver redatuming was correctly performed (according to the standard primary approach). A comparison between figure \ref{fig:50_rtm}b and figure \ref{fig:50_rtm}c shows a distinct difference. Internal multiples indicated by the yellow curved stripes in figures \ref{fig:50_rtm}a and \ref{fig:50_rtm}b are no longer visible in figure \ref{fig:50_rtm}c. Also, the yellow ellipses indicate areas where the removal of internal multiples is clearly visible. Overall, the continuity of the reflectors has improved. Below the vertical RTM images are depth slices of the 3D RTM volume at 5900 m depth, where the internal multiples are present in figures \ref{fig:50_rtm}a and \ref{fig:50_rtm}b, but have been attenuated in figure \ref{fig:50_rtm}c. Based on these results, we conclude that adaptive double-focusing performs well in terms of redatuming and predicting and subtracting internal multiples when applying it to the initial dense acquisition geometry. Note that the image in figure \ref{fig:50_rtm}c appears to have a lower frequency content compared to the images in figures \ref{fig:50_rtm}a and \ref{fig:50_rtm}b. It seems that the removed internal multiple reflections tend to have a high frequency, possibly due to the generation mechanism in the stratified salt. The resulting images in other papers on internal multiple attenuation in the Santos Basin \citep[e.g.][]{griffiths2011applications, cypriano2015impact} confirm this observation.

 In the following, we continue with synthetic data without negative offsets and use source-receiver reciprocity for reconstruction before applying the adaptive double-focusing method.

\subsection{A coarser sail line spacing}

\begin{figure}
	\centering
	\includegraphics[width=12cm]{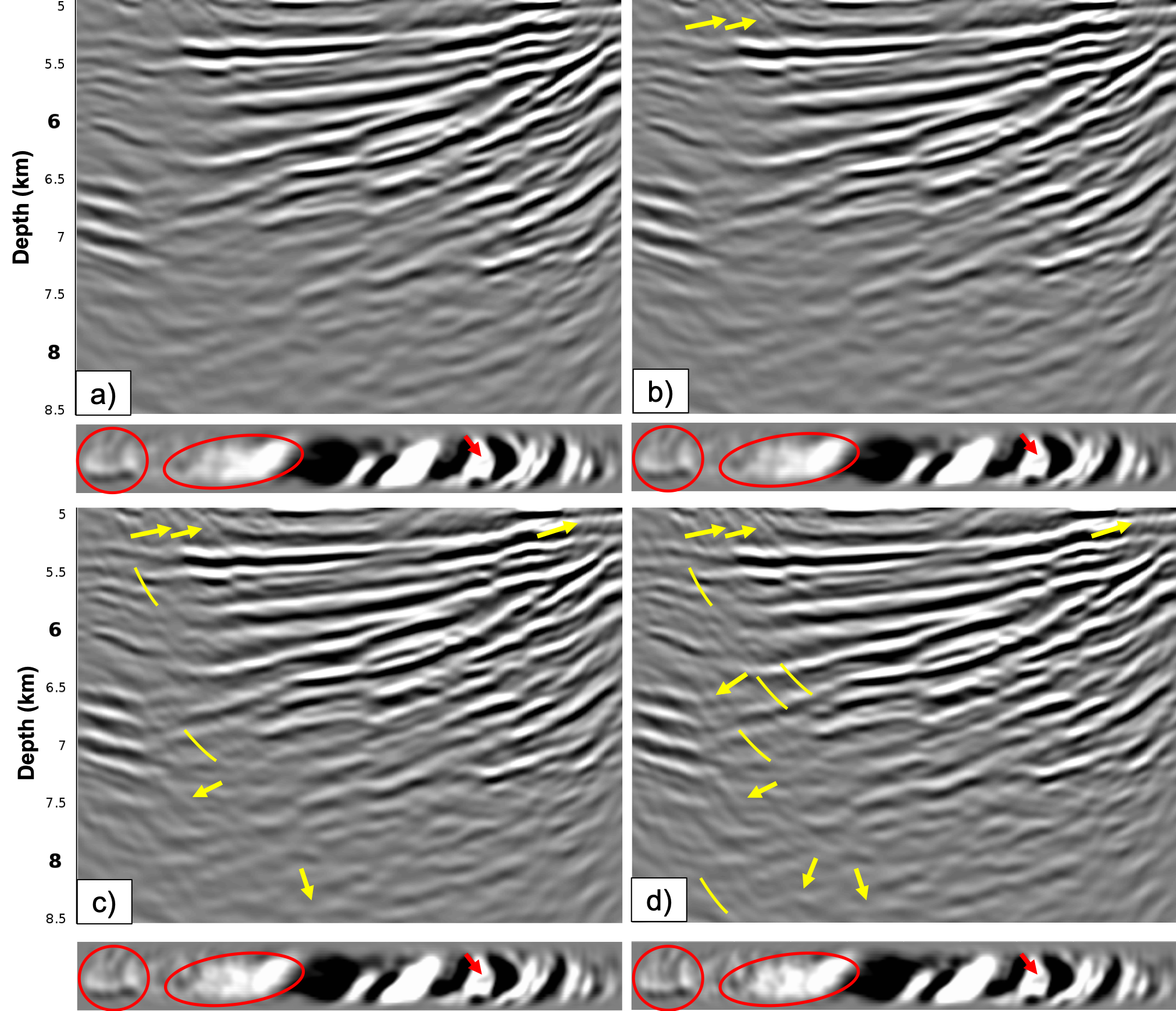}
	\caption{RTM images obtained from the Green's function $\invbreve{G}^{-+}$ after internal multiple prediction and subtraction for a reflection response modeled with: a) 75 m sail line spacing, b) 150 m sail line spacing, c) 300 m sail line spacing and d) 450 m sail line spacing. Below the RTM images are depth slices at 5900 m. The results using 75 m and 150 m sail line spacing are similar, but the image starts to deteriorate when moving to 300 m sail line spacing.
		\label{fig:sail_rtm}}
\end{figure}

Since a sail line spacing of 75 m is not realistic, we study the effect of coarser sail line spacings on the result of our adaptive double-focusing method. Starting from the result in figure \ref{fig:50_rtm}c with 75 m sail line spacing (here figure \ref{fig:sail_rtm}a), we compare RTM images showing the result of adaptive double-focusing when using a sail line spacing of 150 m (figure \ref{fig:sail_rtm}b), 300 m (figure \ref{fig:sail_rtm}c) and finally 450 m (figure \ref{fig:sail_rtm}d). The result obtained from data with a sail line spacing of 150 m looks very similar to the result obtained with 75 m sail line spacing, there are only some minor amplitude differences indicated by the arrows. A more significant difference becomes visible when decimating from 150 m sail line spacing to 300 m sail line spacing. Some internal multiples at the top of the image are no longer predicted and subtracted, probably because the traces that are necessary for the reconstruction of these multiples are missing. The realistic scenario of 450 m sail line spacing shows more internal multiples that could not be predicted and subtracted, now in the deeper part of the image as well. The depth slices confirm these observations: there is little difference between the results obtained with 75 m and 150 m sail line spacing, but internal multiple attenuation becomes less effective when moving to a sail line spacing of 300 m and 450 m. Based on these tests, we conclude that the sail line spacing is a key acquisition parameter that affects our adaptive double-focusing method. Ideally, interpolation from 450 m sail line spacing to 150 m sail line spacing would be applied prior to the field data application in this geological setting. We remark that although we expect that the sail line spacing will also be a key acquisition parameter that affects the result of our adaptive double-focusing methods in other geological settings, the exact spacing at which the result is still acceptable will be different in every setting. In the following synthetic tests, we continue with a sail line spacing of 150 m, thereby assuming that the interpolation from 450 m sail line spacing to 150 m sail line spacing can be carried out correctly.

\subsection{The removal of the near offsets}

\begin{figure}
	\centering
	\includegraphics[width=14cm]{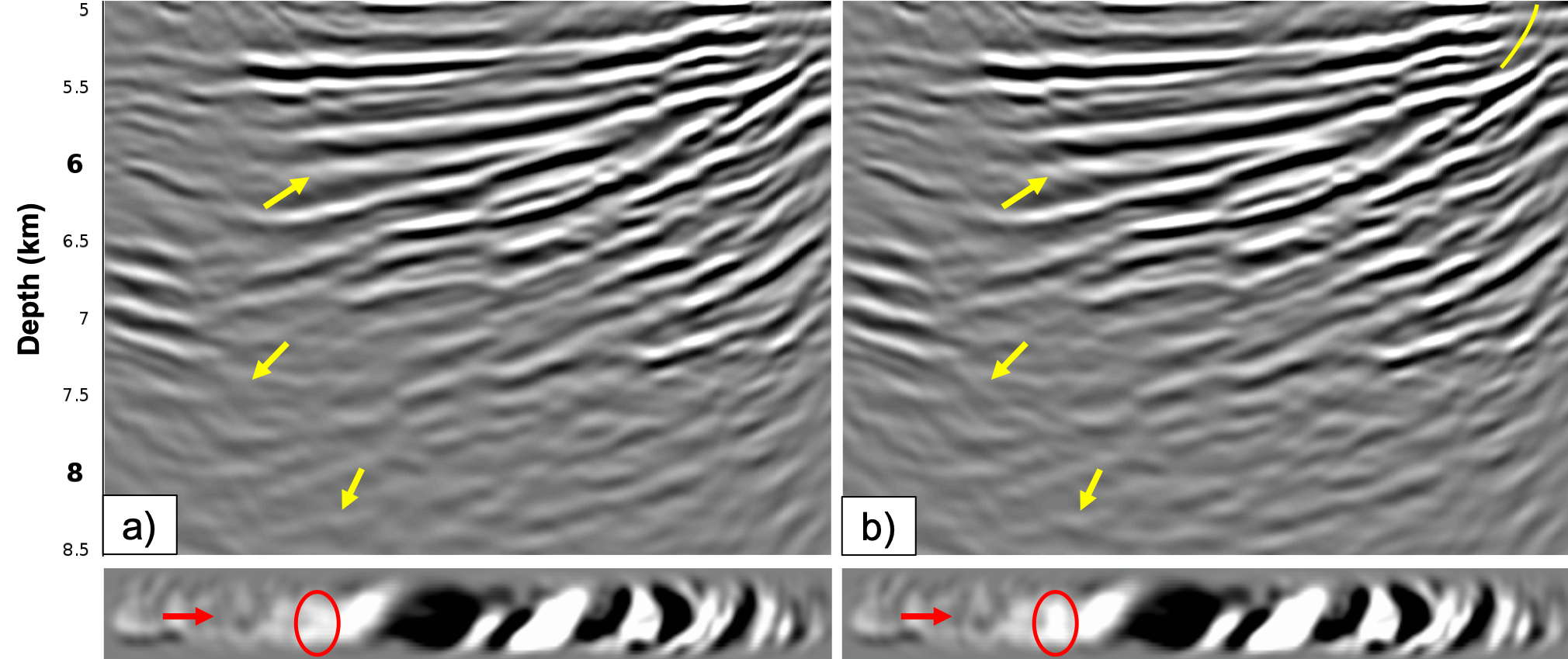}
	\caption{RTM images obtained from the Green's function $\invbreve{G}^{-+}$ after internal multiple prediction and subtraction for a reflection response modeled with: a) near offsets (0-250 m) and b) without near offsets. Below the RTM images are depth slices at 5900 m. Only a slight difference is visible at the arrows, possibly due to the depth of the target zone.
		\label{fig:near_rtm}}
\end{figure}

The responses at near offsets are typically not recorded by streamers, so we study the effect of removing the first 250 m of inline offsets. Figure \ref{fig:near_rtm} shows a comparison of RTM images with and without near offsets, both after internal multiple prediction and subtraction. The removal of the near offsets deteriorates the result somewhat in terms of a few remnant internal multiples (at the ellipse and at the arrows), but not as much as expected. The depth of the first reflector influences how much the near offset responses contribute to the image. Since the water bottom in this setting is very deep (see figure \ref{fig:vel_mod}), most reflections originating from this depth would have simply not been recorded by the first 250 m of receivers. Even though the near offset responses do not have a large effect on the result of internal multiple prediction and removal in this very deep marine setting, we will interpolate the field data for the missing offsets in order to predict as many internal multiples as accurately as possible. 

\subsection{The removal of the far offsets}

\begin{figure}
	\centering
	\includegraphics[width=12cm]{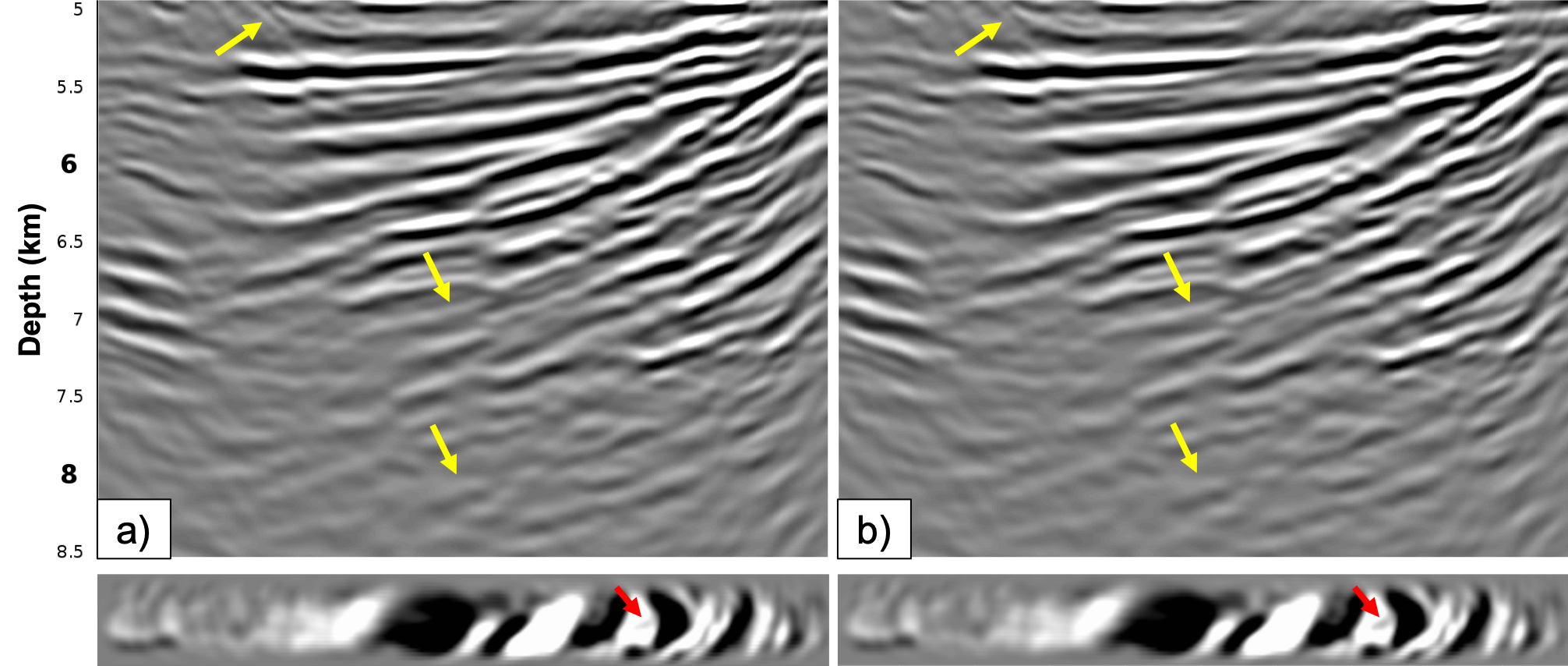}
	\caption{RTM images obtained from the Green's function $\invbreve{G}^{-+}$ after internal multiple prediction and subtraction for a reflection response modeled with: a) far offsets (6250-10000 m) and b) without far offsets. Below the RTM images are depth slices at 5900 m. Only minor differences are visible at the arrows, mostly in terms of amplitude.
		\label{fig:faro_rtm}}
\end{figure}

Next, we assume that we could correctly reconstruct the responses at near offsets and we study the effect of removing the far offsets (the inline offsets 6250-10000 m). Figure \ref{fig:faro_rtm} shows the RTM images of the result of internal multiple prediction and subtraction using adaptive double-focusing, where figure \ref{fig:faro_rtm}a shows the result when including far offsets in the reflection response and figure \ref{fig:faro_rtm}b shows the result when excluding far offsets from the reflection response. Only minor differences are visible in the result, mostly in terms of amplitudes. Surprisingly, the far offsets seem to have little impact on the result of adaptive double-focusing, similar to the near offsets. \cite{verschuur2013seismic} reports that missing offsets have a particularly large effect on multiple prediction methods in a shallow water setting. Since we are in a very deep marine setting, missing offsets seem to only have a minor effect. 

\subsection{The removal of the outer cables}

\begin{figure}
	\centering
	\includegraphics[width=12cm]{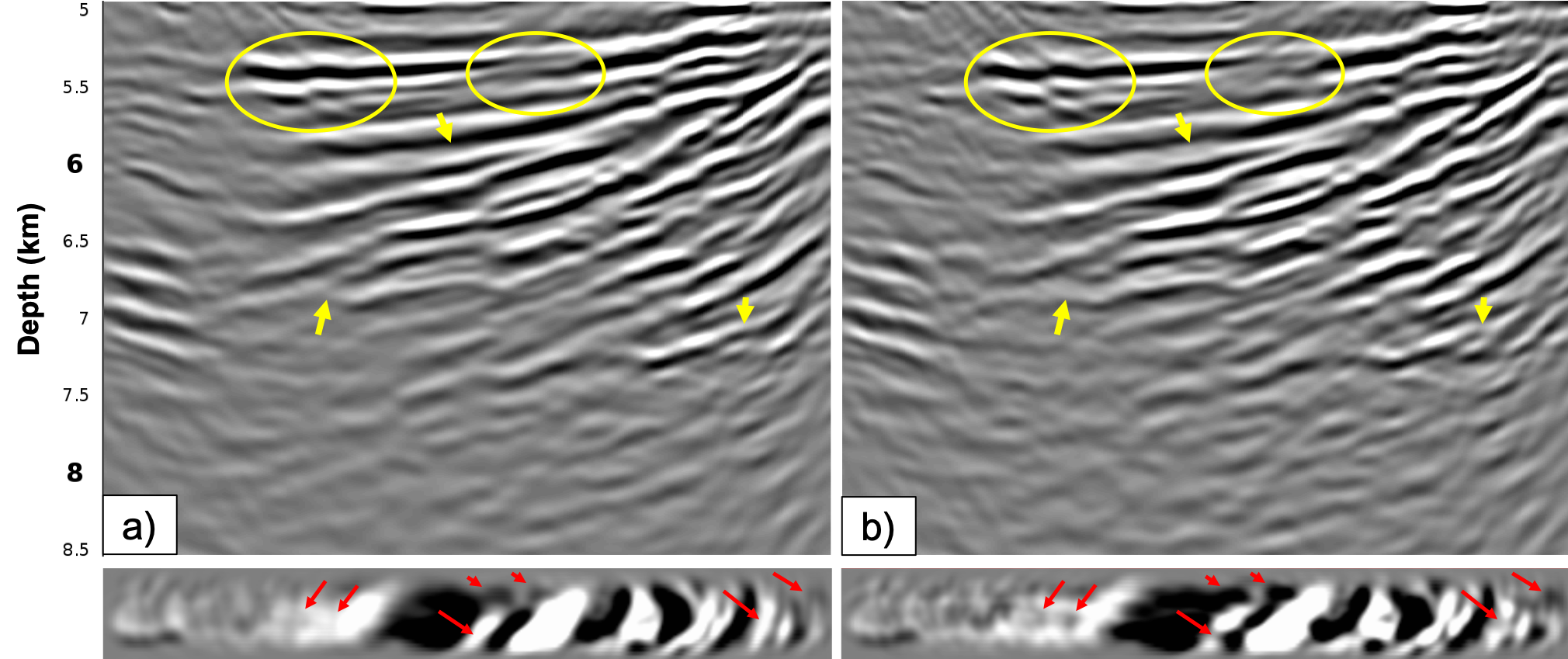}
	\caption{RTM images obtained from the Green's function $\invbreve{G}^{-+}$ after internal multiple prediction and subtraction for a reflection response modeled: a) with outer cables (1.8 km crossline aperture) and b) without outer cables (0.75 km crossline aperture). Below the RTM images are depth slices at 5900 m. The image considerably deteriorates when removing the outer cables, as indicated by the circles and arrows.
		\label{fig:outer_rtm}}
\end{figure}

Lastly, we study the effect of removing the outer cables. Instead of a crossline aperture of 1800 m, as used in the previous tests, we now use a crossline aperture of 750 m. The RTM images in figure \ref{fig:outer_rtm} show that removing the outer cables has a significant effect on the adaptive double-focusing result. The quality of the image in figure \ref{fig:outer_rtm}b has deteriorated and some internal multiples were not predicted and subtracted. Although the missing outer cables have a large effect on the result of adaptive double-focusing, the image in figure \ref{fig:outer_rtm}b is still of acceptable quality. This becomes especially clear when comparing it to the standard redatumed Green's function in figure \ref{fig:50_rtm}b, which is constructed from a dense and wide azimuth grid of sources and receivers at the acquisition surface. Compared to this image, figure \ref{fig:outer_rtm}b still shows a significant reduction in internal multiple energy. This is promising for the field data application, since we cannot compensate for missing outer cables during preprocessing. We remark that the effect of removing the outer cables is expected to become more severe in geological settings with strongly dipping reflectors in the crossline direction. In those cases, the missing outer cables can be a limiting factor that hinders the application of the adaptive double-focusing method. This observation is supported by reports on the performance of similar multiple prediction and removal methods \citep{wang20143d, moore2008general}.

\subsection{The combination of all effects}

\begin{figure}
	\includegraphics[width=15cm]{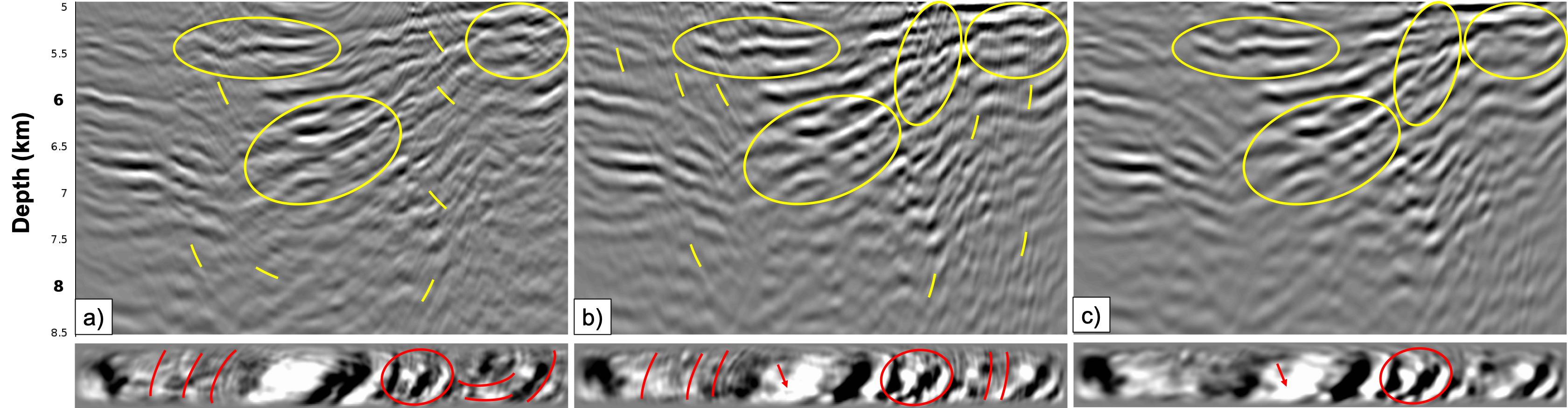}
	\caption{RTM images of the result after applying the adaptive double-focusing method to 3D synthetics with a field data acquistion geometry. The images show: a) the migrated reflection response $\invbreve{R}$, b) the redatumed and migrated Green's function $G_0^{-} * \invbreve{f}_{0}^{+}$ including primaries and internal multiples, and c) the redatumed and migrated Green's function $\invbreve{G}^{-+}$ after prediction and adaptive subtraction of internal multiples. Below the RTM images are depth slices at 5900 m.  
		\label{fig:real_synth_rtm}}
\end{figure}

Although the results of the synthetic tests in the previous sections are encouraging, they do not provide an indication on the feasibility of the application of the adaptive double-focusing method to our field dataset, where all acquisition restrictions are imposed simultaneously. The negative offsets, the near offsets, the far offsets, some sail lines and the outer cables are all missing. Therefore, we model 32 lines of 3D synthetic data based on the acquisition geometry of our narrow azimuth streamer data (figure \ref{fig:grids}b). Next, we reconstruct the negative offsets (by applying source-receiver reciprocity) and the near offsets (by interpolation) and perform interpolation for the sail line spacing (from 450 m to 150 m). Figure \ref{fig:real_synth_rtm} shows the RTM images of the reflection response $\invbreve{R}$, the standard redatumed Green's function $G_0^{-} * \invbreve{f}_{0}^{+}$ with primaries and internal multiples and the redatumed Green's function $\invbreve{G}^{-+}$ after internal multiple prediction and subtraction, zoomed in at the target area. We observe an unexpected difference in illumination when comparing figure \ref{fig:real_synth_rtm}a and figure \ref{fig:real_synth_rtm}b, especially on the right side of the images. Figure \ref{fig:real_synth_rtm}a is constructed by applying an RTM method to the reflection response $\invbreve{R}$, which uses a finite-difference method to back-propagate the wavefield from the acquisition surface to the redatuming level. In contrast, figure \ref{fig:real_synth_rtm}b is constructed by first back-propagating the reflection response $R$ using convolutions with the modeled direct downgoing focusing function, according to $\invbreve{f}_0^{+} * \Psi \mathcal{R} * \invbreve{f}_{0}^+$ (see equation \ref{double}), to obtain $\invbreve{G}_0^{-} * \invbreve{f}_{0}^{+}$, which is subsequently back-propagated from the redatuming level into the target using the RTM method. In principle, back-propagation using a multidimensional convolution is equivalent to back-propagation using an RTM method \citep{esmersoy1988reverse}. However, in practice, these are only equivalent when the same numerical method is used. We use an Eikonal solver to model the direct wave $\invbreve{f}_{0}^{+}$, which is different from the finite-difference method used in the RTM method. As a result, there are slight differences in illumination between the two images.

Next, we evaluate the effectiveness of internal multiple attenuation. A comparison of figures \ref{fig:real_synth_rtm}b and \ref{fig:real_synth_rtm}c shows that the adaptive double-focusing method succeeded in predicting and subtracting internal multiples from the standard redatumed Green's function. Especially inside the yellow ellipses, the internal multiple energy is significantly reduced, resulting in a better continuity of the reflectors. Again, we observe that the internal multiples seem to mainly have a high frequency content. We conclude that the adaptive double-focusing method appears to be sufficiently robust for the prediction and adaptive subtraction of internal multiples from narrow azimuth streamer data in this geological setting.

\section{The 3D field data application}

Based on the results of the synthetic tests, we continue with the field data application. We have 24 lines of narrow azimuth streamer data, acquired using 6 flat streamers with a cable spacing of 150 m and a sail line spacing of 450 m. The length of the cables is 6000 m, covering offsets from 250 to 6250 m. The source and receiver spacing is 50 m in the inline direction. The crossline aperture is 750 m. Deghosting is performed in the f-p domain \citep{wang2013premigration} and a designature filter is obtained from the water bottom reflection. Shot and near offset reconstruction are performed using a partial normal moveout (NMO) correction of traces per common depth point (CDP) \citep[e.g.][]{dragoset2010perspective}. In addition, the data are projected on a regular grid using a $\tau-p$ transform \citep{wang2014fast}. We also remove surface-related multiples, the evanescent wavefield and noise. After preprocessing, we obtain a dataset with a sail line spacing of 150 m. Similar to the synthetic tests, we have chosen the redatuming level just above the base of salt. A smoothed version of the velocity model in figure \ref{fig:vel_mod} and an Eikonal solver are used to model our direct downgoing focusing function $\invbreve{f}_0^{+}$  (including geometerical spreading), which we subsequently convolve with a 30 Hz Ormsby wavelet. Convergence is tracked by computing the $L_2$ norm of the updates of the downgoing focusing function $\invbreve{f}_i^{+}$.

After convolving the individual updates of the wavefields $\invbreve{f}_i^{+}$ and $G_j^{-}$ and migrating them, we obtain the internal multiple predictions in the image domain. Extensive testing shows us that primary reflections are better preserved when subtracting the internal multiple predictions in the image domain instead of in the redatumed domain. Prior to subtraction, the predictions are othogonalized. We use the full curvelet transform (for all scales) and tapered windows of 768 ms by 256 traces.

\begin{figure}
	\centering
	\includegraphics[width=15cm]{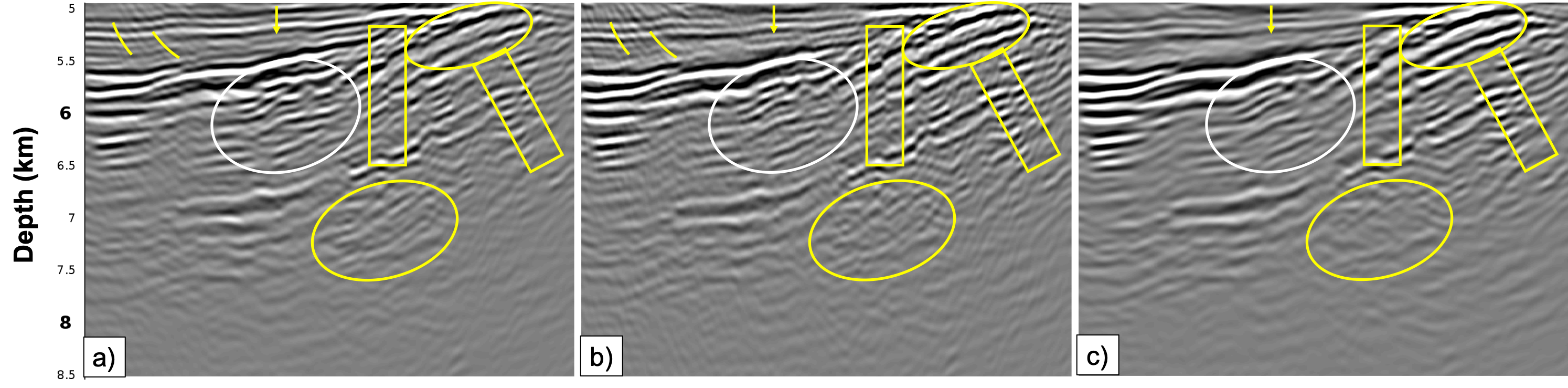}
	\caption{RTM images of the result after applying the adaptive double-focusing method to 3D NAZ streamer data of the Santos Basin, Brazil. The images show: a) the migrated reflection response $\invbreve{R}$, b) the redatumed and migrated Green's function $G_0^{-} * \invbreve{f}_{0}^{+}$ including primaries and internal multiples, and c) the redatumed and migrated upgoing Green's function $\invbreve{G}^{-+}$ after prediction and adaptive subtraction of internal multiples. Below the RTM images are depth slices at 5900 m. Internal multiples were predicted and subtracted, resulting in an improved image of the target area.}
	\label{fig:field_rtm}
\end{figure}

\begin{figure}
	\centering
	\includegraphics[width=13cm]{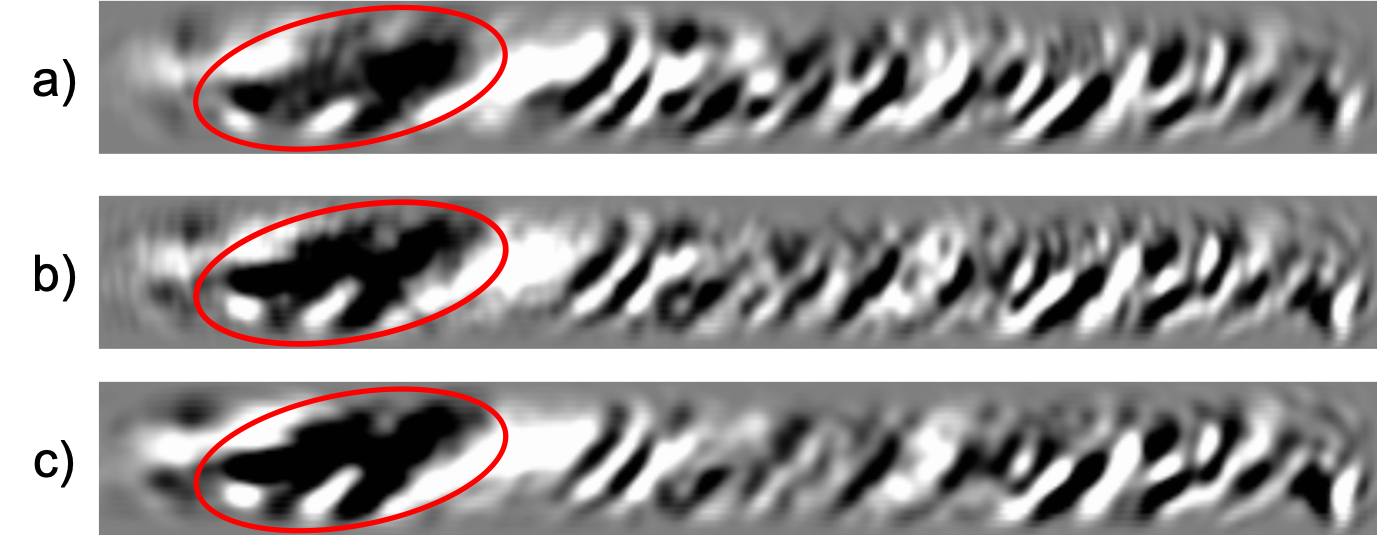}
	\caption{Depth slices corresponding to the RTM images in figure \protect\ref{fig:field_rtm} at 5900 m. The slices show: a) the migrated reflection response $\invbreve{R}$, b) the redatumed and migrated Green's function $G_0^{-} * \invbreve{f}_{0}^{+}$ including primaries and internal multiples, and c) the redatumed and migrated upgoing Green's function $\invbreve{G}^{-+}$ after prediction and adaptive subtraction of internal multiples.}
	\label{fig:field_depth}
\end{figure}

Figures \ref{fig:field_rtm} and \ref{fig:field_depth} show the result of applying the adaptive double focusing method to predict and adaptively subtract internal multiples. First, we observe that there is still a slight illumination difference between the RTM migrated image of the reflection response in figure \ref{fig:field_rtm}a and the image of the RTM migrated redatumed reflection response in figure \ref{fig:field_rtm}b (see the yellow circle in the top right), which has been explained above. However, the difference is not as pronounced as in figure \ref{fig:real_synth_rtm}. Second, a clear difference is visible between figures \ref{fig:field_rtm}b and figure \ref{fig:field_rtm}c, indicating that the adaptive double-focusing method succeeded in predicting and adaptively subtracting events which are likely internal multiples. The arrow, the ellipses and the lower yellow circle indicate areas in which events were attenuated, while the yellow boxes indicate what we interpret to be an improvement in fault definition. The white circle shows conflicting events that are being resolved, but it also shows events that we believe to be remnant internal multiples. These were most likely generated by the base of salt, which is not part of our overburden. In order to also remove these events and to see if there is perhaps a hidden structure, the redatuming level could be placed just below the base of salt. Similar to what we observed in the synthetic data, the image after internal multiple attenuation seems to have a lower frequency content. The RTM depth slices in figure \ref{fig:field_depth} demonstrate the attenuation of events in the form of ellipses. This example shows that the adaptive double-focusing method is sufficiently robust for the application to our 3D narrow azimuth streamer dataset. Naturally, this result can be improved by the use of more lines of data.

\section{Discussion and Conclusion}

In this paper, we identified the key acquisition parameters that affect the application of our adaptive Marchenko internal multiple attenuation method to narrow azimuth streamer data. Tests on 3D synthetic data evaluated the effect of removing sail lines, near offsets, far offsets and the outer cables. The results of these tests show that the aperture in the crossline direction and the sail line spacing have the strongest effect on the quality of the result. Typically, the sail line spacing can be interpolated, but the aperture in the crossline direction can possibly be a limiting factor for our method. Surprisingly, the missing near offsets and the far offsets only had a modest effect on the result of our method, possibly due to the very deep target area. In addition, we found that the responses at the negative offsets and the near offsets could be accurately reconstructed. We remark that these tests are only valid for this particular dataset, but they give an impression of the possibilities and limitations of the adaptive Marchenko double-focusing method. For an Ocean Bottom Node (OBN) acquisition geometry, these tests imply that our method will be most sensitive to the node separation (especially in the direction of the strongest geological variation). 

Based on the decimation tests, we defined an interpolation strategy that was first tested on a realistic synthetic dataset. We reconstructed the negative offsets and the near offsets, and interpolated the sail line spacing from 450 m to 150 m. When applying the 3D adaptive double-focusing method, an important aspect that was not visible in earlier 2D applications became visible, thereby showing that the extension of a method from 2D to 3D is not always trivial. In 3D, when using an Eikonal solver for the modeling of the direct wave and a finite-difference-based RTM method, a slight difference in illumination between the RTM image of the reflection response and the RTM image of the redatumed response occurs. Nevertheless, the double-focusing method predicted and adaptively subtracted internal multiples, thereby improving the image of the target area. 

Next, we applied the adaptive double-focusing method to 24 lines of narrow azimuth streamer data. We reconstructed the negative and the near offsets and interpolated the sail line spacing. We interpret that internal multiples were predicted and adaptively subtracted, which resulted in an improved geological interpretation of the target area. Therefore, we conclude that 3D Marchenko internal multiple attenuation using an adaptive double-focusing method is sufficiently robust for the application to narrow azimuth streamer data in a deep marine setting, provided that there is sufficient aperture in the crossline direction and that the sail lines are interpolated.

Note that redatuming is optional for Marchenko methods. The adaptive double-focusing method used in this paper includes source-receiver redatuming, which is particularly useful when the aim is to attenuate internal multiples in the target area and at the same time reducing the data volume for a next processing step (for example, a target-oriented full waveform inversion). In contrast, when the aim is only to attenuate internal multiples as part of a larger workflow, the adaptive double-focusing method might not be the Marchenko method of choice. A direct quality check of the input common source gathers and the redatumed common source gathers is not possible, which is a disadvantage in a general processing workflow. In addition, a quality check on the resulting images is only possible when the same numerical method is used to obtain the direct wave for the Marchenko method  as for the migration of the original data. Therefore, for the purpose of internal multiple elimination only, we propose the use of other Marchenko methods that do not include redatuming and thus allow for an easier quality check. An example is the adaptive overburden elimination method \citep{van2016adaptive}, as shown in papers by \cite{pereira2018efficient}, \cite{krueger2018internal} and \cite{pereira2019internal}. A modified version of the adaptive double-focusing method is presented by \cite{staring2019adaptive}. Other alternatives are the Marchenko multiple elimination scheme \citep{zhang2018marchenko} and the primary-only method proposed by \cite{meles2016reconstructing}. 

The Marchenko method used in this paper is acoustic. The synthetic data are acoustic, but naturally the field data are elastic. A suggestion for further research is to evaluate the effect of the presence of mode conversions on the acoustic Marchenko method in this geological setting, as was done by \cite{reinicke2019do} for the offshore Middle East. By applying  an acoustic Marchenko method to elastic synthetic data, and comparing it to the result of applying an acoustic Marchenko method to acoustic synthetic data, \cite{reinicke2019do} evaluated whether the acoustic approximation is valid for structural imaging in the region. They concluded that the acoustic approximation may be sufficient when used for structural imaging in 1.5D geological settings.

\section*{acknowledgements}
This research was performed in the framework of the project 'Marchenko imaging and monitoring of geophysical reflection data', which is part of the Dutch Open Technology Programme with project number 13939 and is financially supported by NWO Domain Applied and Engineering Sciences. The research of K. Wapenaar has received funding from the European Research Council (ERC) under the European Union's Horizon 2020 research and innovation programme (grant agreement no. 742703). We thank Roberto Pereira, Peter Mesdag and Adel Khalil from CGG for the close collaboration and for providing the field data. Also, we would like to thank two anonymous reviewers, Lele Zhang, Giovanni Meles and Jan Thorbecke for valuable discussions.

\section*{Data Availability Statement}
Data sharing is not applicable to this article.

\bibliography{thesis}

\end{document}